\input{aipcheck}
\documentclass[
    ,final            
  ]
  {aipproc}
\layoutstyle{8x11double}

\begin{document}

\def\eg{{\it e.g.,}}
\def\etal{{\it et al.}}
\def\ie{{\it i.e.,}}

\title{Comparison of CORSIKA and COSMOS simulations}
\author{Soonyoung Roh}{
  address={Department of Astronomy and Space Science, Chungnam National University, Daejeon 305-764, Korea}
}
\author{Jihee Kim}{
  address={Department of Astronomy and Space Science, Chungnam National University, Daejeon 305-764, Korea}
}
\author{Katsuaki Kasahara}{
  address={Institute for Cosmic Ray Research, University of Tokyo, Chiba 277-8582, Japan}
}
\author{Eiji Kido}{
  address={Institute for Cosmic Ray Research, University of Tokyo, Chiba 277-8582, Japan}
}
\author{Akimichi Taketa}{
  address={Center for High Energy geophysics Research, Earthquake Research Institute, University of Tokyo, Chiba 277-8582, Japan}
}
\author{Dongsu Ryu}{
  address={Department of Astronomy and Space Science, Chungnam National University, Daejeon 305-764, Korea}
}
\author{Hyesung Kang}{
  address={Department of Earth Sciences, Pusan National University, Pusan 609-735, Korea}
}

\begin{abstract}
Ultra-high-energy cosmic rays (UHECRs) refer to cosmic rays with energy above 10$^{18}$eV. UHECR experiments utilize simulations of extensive air shower to estimate the properties of UHECRs. The Telescope Array (TA) experiment employs the Monte Carlo codes of CORSIKA and COSMOS to obtain EAS simulations. In this paper, we compare the results of the simulations obtained from CORSIKA and COSMOS and report differences between them in terms of the longitudinal distribution, Xmax-value, calorimetric energy, and energy spectrum at ground.
\end{abstract}

\maketitle
\section{I. INTRODUCTION}

Ultra-high-energy cosmic rays (UHECRs) with energy greater than 10$^{18}$eV arrive at Earth from space. Experiments to detect UHECRs utilize extensive air shower (EAS) to estimate their energy, composition and arrival direction. In an EAS, the cascade of interactions, induced by a UHECR in the upper atmosphere, results in a very large number of secondary particles: of order 10$^{12}$ particles for a primary particle with energy 10$^{19}$eV.

Monte Carlo codes have been used to simulate EAS. The codes reproduce cascades of particles initiated by the interaction between primary UHECRs and atmospheric nuclei. EAS simulations give the spatial, temporal, energy, and angular distributions of secondary, air shower particles. To extract the information of primary particles in experiments, that is, the energy, composition and arrival direction, the measured quantities of EAS need to be compared with those from simulations. Hence, the accurate reproduction of EAS is an essential part of UHECR experiments.

CORSIKA \cite{D.Heck} and COSMOS \cite{K.Kasahara} are among the Monte Carlo codes. Most experiments including the recent one at the Pierre Auger Cosmic Ray Observatory have relied simulations from CORSIKA. On the other hand, the Telescope Array (TA) experiment \cite{TA} has employed both CORSIKA and COSMOS, to cross-check the simulations from the codes. In this paper, we compare CORSIKA and COSMOS simulations by examining the differences in the quantities such as the longitudinal distribution, Xmax-value, calorimetric energy, and energy spectrum at ground in EAS.

\section{II. SIMULATIONS}

We employed the most recent versions of codes: Version 7.54 for COSMOS and Version 6960 for CORSIKA. Each code has an option to set interaction models; different interaction models result in somewhat different results. We chose the following interaction models: 1) for CORSIKA, QGSJETII-03 \cite{qgsjet-II} for high-energy (above 80 GeV) hadronic interactions, FLUKA (v.2008.3c) for low-energy (below 80 GeV) hadronic interactions, and EGS4 for electromagnetic interactions, 2) for COSMOS, QGSJETII-03 for high-energy (above 80 GeV) hadronic interactions, PHITS \cite{phits} and JAM \cite{jam} for low-energy (below 80 GeV) hadronic interactions, and Tasi's and Nelson's formula for electromagnetic interactions. Please see references for CORSIKA and COSMOS for details.

The above interaction models are the default, so most-widely used models. In this paper, we intend to compare the results of most-commonly employed CORSIKA and COSMOS simulations. We note that even with the same interaction models, the results of CORSIKA and COSMOS simulations could be different, because of the differences in handling the development of EAS. Comparisons of CORSIKA and COSMOS simulations with different interaction models will be reported in an upcoming journal paper.

We adopted exactly same parameters for CORSIKA and COSMOS simulations. For thinning parameter, 10$^{-7}$ was used (Hillas thinning algorithm) \cite{Hillas1971}. The ground level was located at 875 g/cm$^{2}$ (1430 m), and the atmospheric depth and the Earth magnetic field suitable for the TA site were applied.

We generated air shower events with cosine of the zenith angle of 1, 0.95, 0.9, $\cdots$, 0.7 and primary energy of 10$^{18.5}$, 10$^{18.75}$, 10$^{19}$, $\cdots$, 10$^{20.25}$ eV for proton and iron primaries. Altogether, about 10,000 showers were generated with each of CORSIKA and COSMOS. The results below are based on some of the shower simulations.

\begin{figure}
\includegraphics[width=79mm]{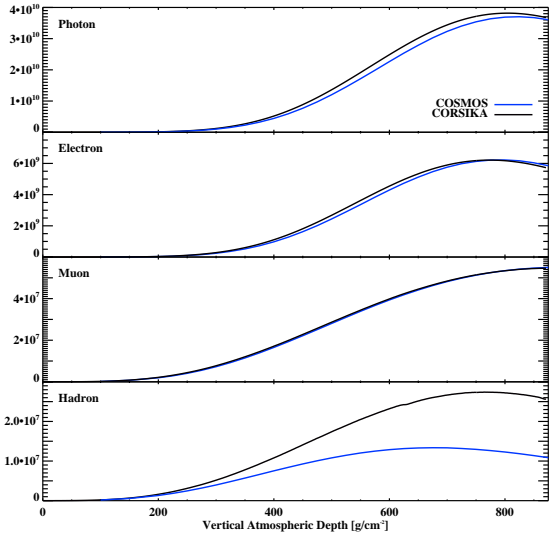}
\caption{Longitudinal development of proton EAS with the primary energy 10$^{19}$eV and the zenith angle 0 degree. Panels show the numbers of photons, electrons, muons, and hadrons as a function of atmospheric depth. Black lines are the CORSIKA results, while blue lines are the COSMOS results. The lines are averages of 50 shower events.}
\end{figure}

\section{III. COMPARISONS}

\subsubsection{1. LONGITUDINAL DISTRIBUTION}
Figure 1 shows a typical longitudinal development of air showers obtained with CORSIKA and COSMOS. Overall, the numbers of secondary particles are predicted to be larger with CORSIKA than with COSMOS. There are noticeable differences. For instance, the maximum difference in the photon number reaches up to $\sim 10 \%$. The difference in the hadron number is much larger, although the number itself is much smaller than those of other particles. On the other hand, the agreement in the electron and muon numbers is in a acceptable level; the maximum difference in the electron number is $\sim 5 \%$ or less.

\subsubsection{2. Xmax}
Xmax is the atmospheric depth of the shower maximum, specifically the maximum of the electron distribution. It is one of most important quantities in UHECR experiments; it is mainly used to determine the composition. We employed the Geisser-Hillas function \cite{gh1977} to determine the longitudinal development and Xmax, which has been widely employed in other studies. Figure 2 shows Xmax for both proton and iron primaries from COSMOS and the CORISKA simulations. In general, iron-initiated air showers penetrate less than proton-initiated air showers, so Xmax for iron primaries is less than that for proton primaries. And Xmax increases with the energy of primaries. The agreement in COSMOS and the CORSIKA simulations is reasonably good; the difference in Xmax is only a few percent at most.

\begin{figure}
\includegraphics[width=79mm]{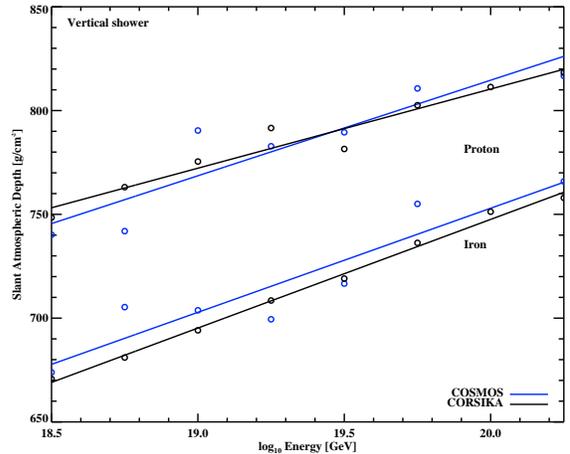}
\caption{Xmax as a function of the primary energy. The results shown are for vertical showers with zenith angle 0 degree. The upper part is for proton primaries, and the lower part is for iron primaries. Black dots are the CORSIKA results, while blue dots are the COSMOS results. Those are averages of 50 shower events. Lines are least chi-square fits of points.}
\end{figure}

\subsubsection{3. CALORIMETRIC ENERGY}
As EASs develops, a part of the energy of primary particles ($E_0$) is deposited into air molecules and eventually radiated as fluorescence lights. But a fraction of the energy is carried away by secondary particles, not contributing to fluorescence lights. A correction for the so-called missing energy must be applied to the measurement of the calorimetric energy ($E_{cal}$), in order to correctly determine the primary energy, $E_0$, from observation of fluorescence lights. Here, we calculated the missing energy by following the prescription described by \cite{barbosa2004}, and so the calorimetric energy. Figure 3 shows the resulting $E_{cal}$ for proton primaries from CORSIKA and COSMOS simulations. Our result indicates that $E_{cal}$ from COSMOS is $\sim 2 \%$ larger than that from CORSIKA. This implies that the primary energy estimated with fluorescence detector would be $\sim 2 \%$ smaller when COSMOS simulations are used.

\begin{figure}
\includegraphics[width=79mm]{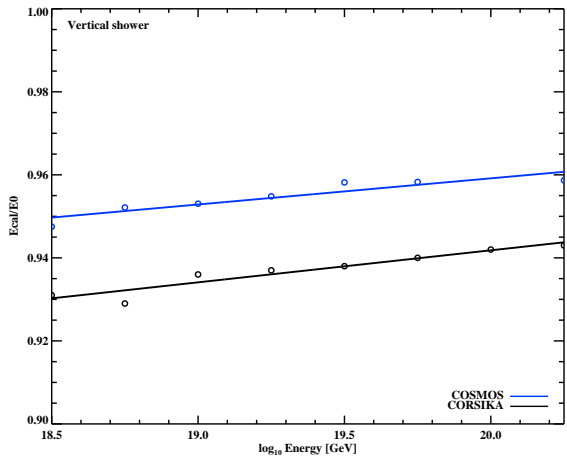}
\caption{Calorimetric energy as a function of the primary energy for proton primaries. The results shown are for vertical showers with zenith angle 0 degree. Black dots are the CORSIKA results, while blue dots are the COSMOS results. Those are averages of 50 shower events. Lines are least chi-square fits of points.}
\end{figure}

\subsubsection{4. ENERGY DISTRIBUTION AT THE GROUND}

As a consequence of EAS, a number of secondary particles arrive at ground. In experiments, those particles are registered by surface detectors and used to estimate the energy and arrival direction of the primary particles. In Figures 4, 5, and 6, we compare the energy (rest-mass energy no included) distribution of photons, electrons, and muons reached at ground from CORSIKA and COSMOS simulations. Particles in the core are included. The figure indicates a good agreement between the CORSIKA and COSMOS results; the typical difference is $\sim 3 \%$ or so. Although not shown here, the difference in the distribution of hadrons is much larger. But again, the number and energy of hadrons are much smaller than those of other particles.
\begin{figure}[htbp]
\includegraphics[width=55mm,angle=-90]{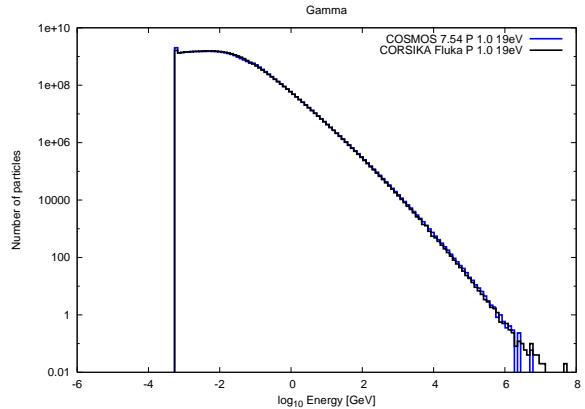}
\caption{Energy (rest-mass energy no included) distribution of photons at ground over the entire ground for proton EAS with the primary energy 10$^{19}$eV and the zenith angle 0 degree. Black lines are the CORSIKA results, while blue lines are the COSMOS results. The lines are averages of 50 shower events.}
\end{figure}

\begin{figure}[htbp]
\includegraphics[width=55mm,angle=-90]{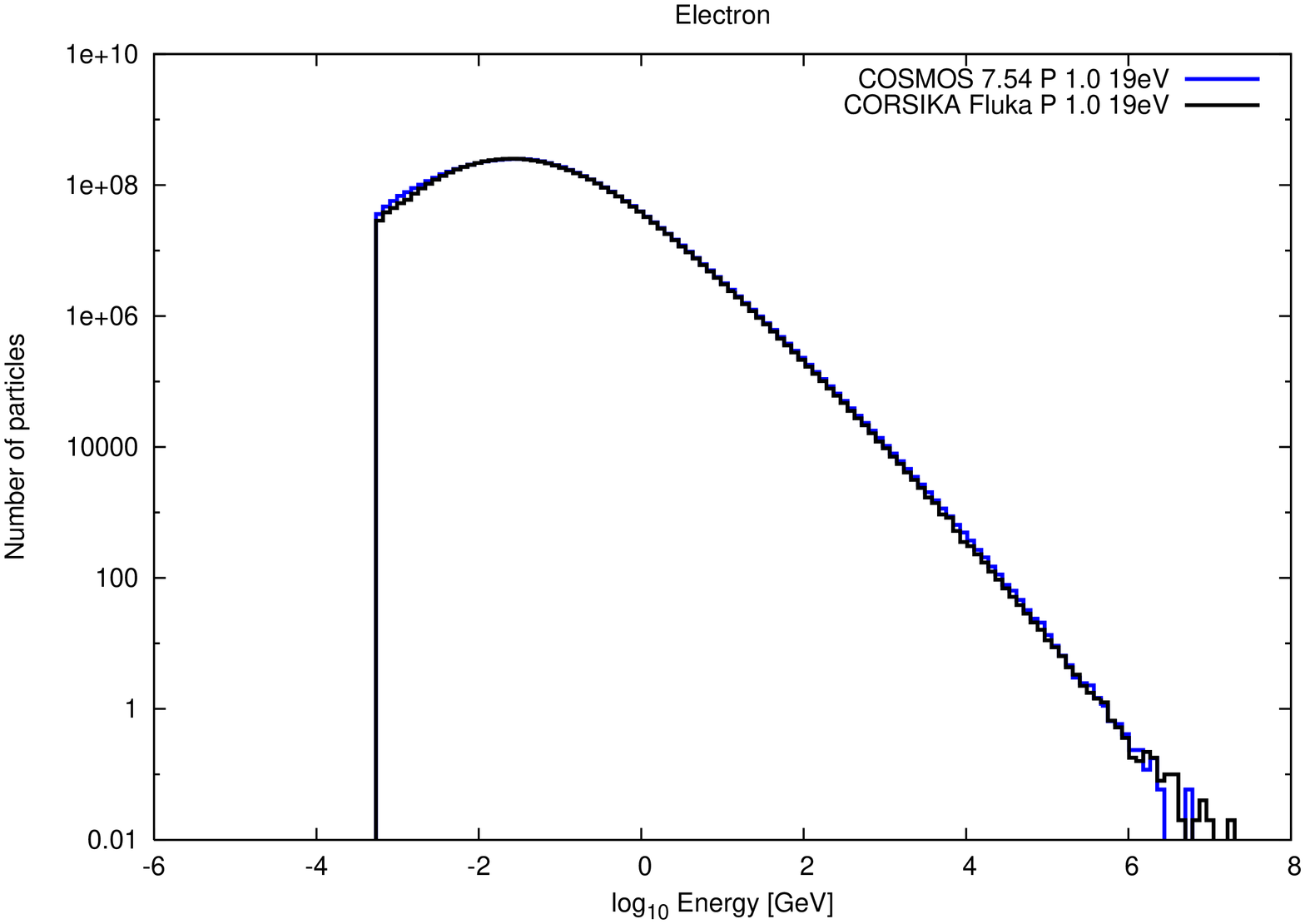}
\caption{Energy (rest-mass energy no included) distribution of electrons at ground over the entire ground for proton EAS with the primary energy 10$^{19}$eV and the zenith angle 0 degree. Black lines are the CORSIKA results, while blue lines are the COSMOS results. The lines are averages of 50 shower events.}
\end{figure}

\begin{figure}[htbp]
\includegraphics[width=55mm,angle=-90]{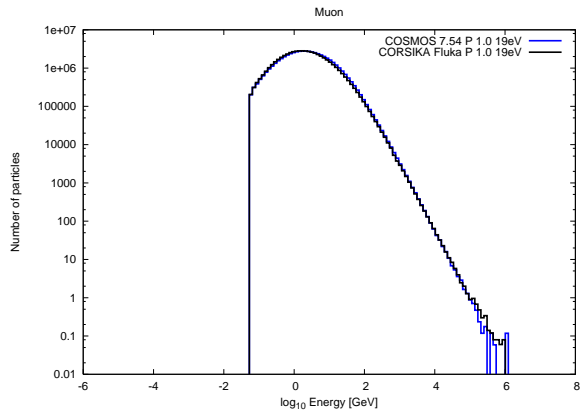}
\caption{Energy (rest-mass energy no included) distribution of muons at ground over the entire ground for proton EAS with the primary energy 10$^{19}$eV and the zenith angle 0 degree. Black lines are the CORSIKA results, while blue lines are the COSMOS results. The lines are averages of 50 shower events.}
\end{figure}

\section{IV. DISCUSSION}

Monte Carlo codes of CORSIKA and COSMOS are currently used to analyze the data of the TA experiment. In this paper, we compared simulations of EAS using CORSIKA and COSMOS codes and quantified the differences in the simulations. For the longitudinal distribution of photons, electrons, and muons, we found the maximum difference of $\sim 10 \%$. For Xmax and calorimetric energy, the difference is typically a few percent. The difference in the energy distribution of photons, electrons, and muons at ground is again typically a few percent. This implies that we should expect an uncertainty of a few percent in the estimation of the primary energy in UHECR experiments including the TA experiment, solely due to the uncertainty in EAS simulations.

Finally, it is worthwhile to mention that there is a large difference in the production of secondary hadrons in the CORSIKA and COSMOS codes. We found that the production is rather sensitive to the low-energy (below 80 GeV) hadronic interaction model. This may indicate that low-energy hadronic interaction models in Monte Carlo codes for EAS need to be further investigated.

\begin{theacknowledgments}

The work was supported by the National Research Foundation of Korea through grant 2007-0093860 and Grant-in-Aid for Scientific Research on Priority Areas (Highest Cosmic Rays: 15077205) of MEXT, Japan.

\end{theacknowledgments}

\bibliographystyle{aipproc}   
\bibliography{sample}

\IfFileExists{\jobname.bbl}{}
 {\typeout{}
  \typeout{******************************************}
  \typeout{** Please run "bibtex \jobname" to optain}
  \typeout{** the bibliography and then re-run LaTeX}
  \typeout{** twice to fix the references!}
  \typeout{******************************************}
  \typeout{}
 }

\end{document}